\documentclass[12pt, twocolumn]{IEEEtran}
\usepackage{colortbl}
\usepackage{color}
\usepackage{graphicx}
\usepackage{float} 
\usepackage{subfigure} 
\usepackage{booktabs}
\ifCLASSINFOpdf
\else
\fi
\begin{document}
\title{Transient Analysis for Resonant Beam \\ Charging and Communication}
%
%
%

\author{Jie~Zhou,
        Mingliang~Xiong,
        Mingqing~Liu,
        Qingwen~Liu,~\IEEEmembership{Senior Member,~IEEE,}
        and~Shengli~Zhou,~\IEEEmembership{Fellow,~IEEE}
\thanks{Copyright (c) 20xx IEEE. Personal use of this material is permitted. However, permission to use this material for any other purposes must be obtained from the IEEE by sending a request to pubs-permissions@ieee.org.

Manuscript received Apr 24, 2021; revised Jun 17, 2021; accepted Jun 29, 2021. The work was supported by the National Key R\&D Program of China under Grant 2020YFB2103900 and Grant 2020YFB2103902. It was also supported by the National Natural Science Foundation of China under Grant 61771344 and Grant 62071334. (Corresponding author: Qingwen Liu.) 

Jie Zhou, Mingliang Xiong, Mingqing Liu, and Qingwen Liu are with Department of Computer Science and Technology, Tongji University, Shanghai, People's Republic of China (e-mail: jzhou@tongji.edu.cn, xiongml@tongji.edu.cn, clare@tongji.edu.cn, qliu@tongji.edu.cn). 
Shengli Zhou is with Department of Electrical and Computer Engineering, University of Connecticut, Storrs, CT, 06250 (e-mail: shengli.zhou@uconn.edu.)}

}

\maketitle

\begin{abstract}
High communication speed and sufficient energy supply are the directions of technological development.
Energy and information available anywhere and anytime has always been people's good wishes.
On this basis, resonant beam system (RBS) has demonstrated its unique superiority in meeting the needs for energy and communication.
The previous work has mostly focused on the analysis of  charging performance of RBS and its steady-state characteristics.
In order to analyze the communication performance of RBS more thoroughly, we propose a resonant beam charging and communication (RBCC) system  and use the equivalent circuit analysis method to conduct transient analysis on it. 
The equivalent circuit reveals the dynamic establishment process of the resonant beam from scratch, 
which facilitates the analysis of the relaxation oscillation 
process and
a deeper understanding of the energy transmission and communication performance. 
In addition, we explore the energy transmission and communication performance of the RBCC under different energy allocation strategies.
\end{abstract}

\begin{IEEEkeywords}
SWIPT, equivalent circuit, transient analysis.
\end{IEEEkeywords}

%
\IEEEpeerreviewmaketitle

\section{Introduction}\label{sec:INTRO}
\IEEEPARstart{W}{ith} the widespread application of Internet of Thing (IoT) devices, the industry estimates that there are tens or even hundreds of billions of IoT devices connected to wireless networks to provide ubiquitous connections \cite{lee2015internet}. 
In addition to logistics factories~\cite{8375735}, smart homes~\cite{8661504}, smart cities~\cite{kolozali2018observing} and other places with plenty of electricity, IoT has played a huge role in difficult and even dangerous places, such as earthquakes~\cite{8447195}, deserts, and landslides~\cite{8895751}. Ensuring and extending battery life is one of the most important considerations for IoT devices~\cite{7954594,8443433}. For medical devices such as pacemakers, the life of the device can mean the difference between life and death~\cite{sametinger2015security}.
Although low-power wireless communication technologies (such as LoRa \cite{croce2019lora}, NB-IoT \cite{8700669,9210822}, etc.)  enable the battery life of IoT devices to several years, the labor cost of replacing batteries with a huge number of IoT devices is unacceptable~\cite{ma2019sensing}. 
Therefore, in the future, it is hoped that IoT terminals will be as passive as possible, or the battery life will be infinite.
Without earth-shaking innovations in the battery field~\cite{li2019wirelessly}, the battery life is difficult to satisfy users.
Thus, wireless power transfer (WPT) or wireless charging technology has become a good choice.

\begin{figure}[t]
    \centering
    \includegraphics[width=3.2in]{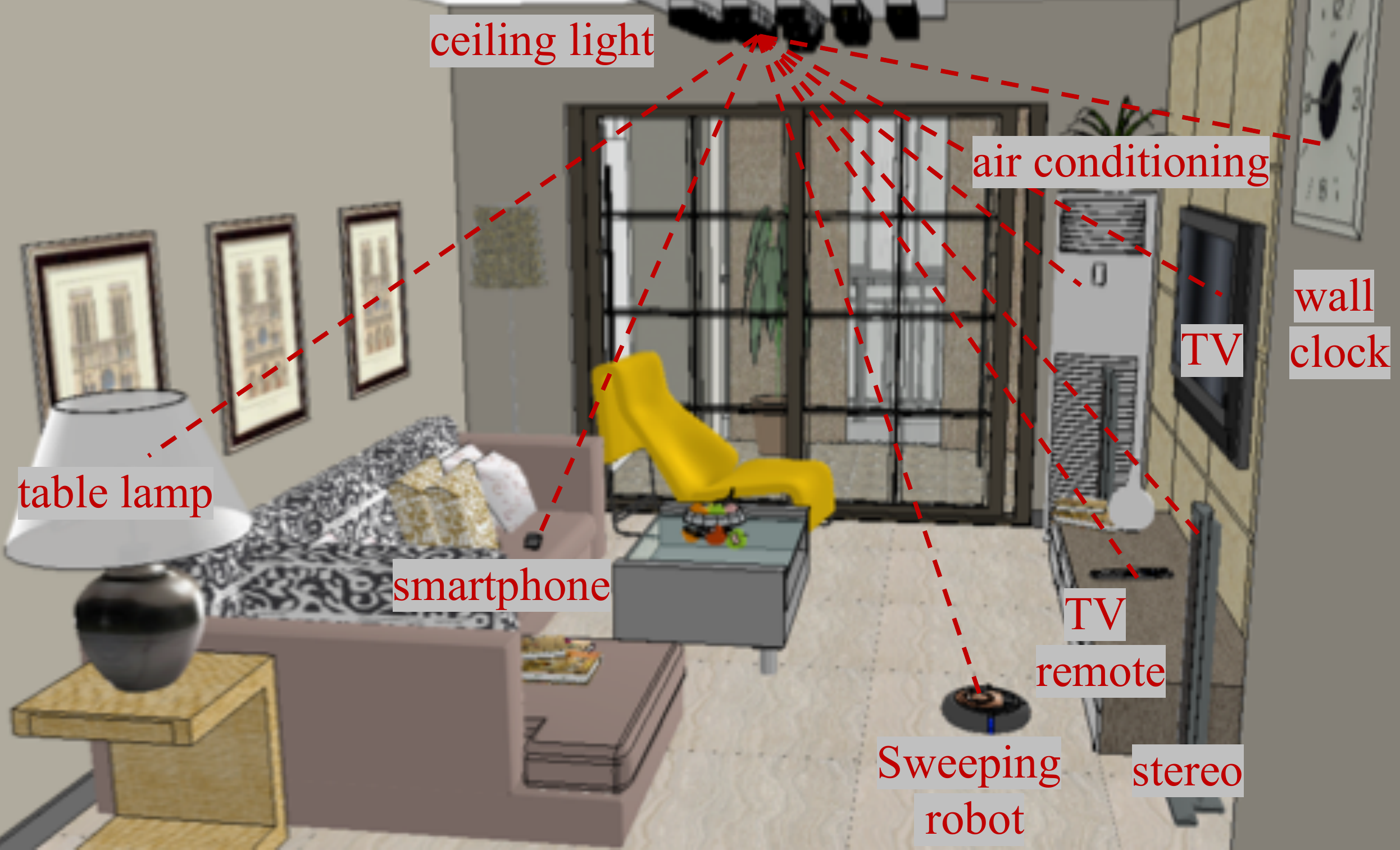}
    \caption{A typical application scenario of RBS.}
    \label{f:intro}
\end{figure}

The wireless power transmission technology allows the charger to get rid of the limitation of the line and realize the complete separation of electrical appliances and power. In terms of safety and flexibility, WPT shows better advantages than traditional chargers.
Three main WPT technologies are investigated in \cite{ho2011comparative, beh2012automated, sambo2014survey}.
Inductive coupling~\cite{ho2011comparative} is safe and easy to implement. However, it is limited by a short charging distance from a few millimeters to a few centimeters, which is only suitable for contact charging devices such as toothbrushes.
Magnetic resonance  coupling~\cite{beh2012automated} has high charging efficiency. However, it is limited by a short charging distance and a large coil size, and is suitable for household appliances such as TVs. 
Electromagnetic (EM)  radiation~\cite{sambo2014survey} has a long effective charging distance, while its charging efficiency is low and it is not safe when exposed to high EM power density. So EM is only suitable for low-power devices such as sensors. 
In short, these traditional WPT technologies provide excellent wireless charging capabilities for some application scenarios, but have their own limitations in other areas.
A safe technology which provides sufficient power for IoT and mobile devices (for example, smart sensors, smart phones, laptops, etc.) at long distances is strongly needed. It is usually necessary to provide watt-level power at meter-level distances.

At the same time, communication is also a very important part for IoT equipments.
The communication capabilities of IoT devices allow them to be connected as a whole system and form a unified network.
The wireless connection is very complicated, and the dense equipment deployment further complicates the operation~\cite{7582463}. Therefore, achieving information flow among devices, infrastructure, cloud, and applications is a large IoT challenge. For billions of devices, radio channel congestion is a problem that will only get worse~\cite{9123361}.
Thus, we are thinking about finding a solution which is outside the radio domain, and at least alleviating the communication pressure of IoT devices.

Instead of radio, resonant beam system (RBS) \cite{fang2018fair, zhang2018adaptive} has demonstrated its unique advantages in meeting the energy and communication requirements of IoT devices.
RBS can be used as a wireless energy transmission method that can realize long-distance, high-power, safe, and mobile charging. 
The main part of RBS is a space-separated resonant cavity, which is mainly composed of a transmitter and a receiver \cite{fang2018fair}.
At the same time, the RBS has a self-alignment function, and no additional measures are required to ensure the alignment between the receiver and the transmitter. Therefore, RBS has a high degree of mobility.
Besides, RBS can achieve a power transmission exceeding 5 W and a distance transmission  exceeding 8 meters~\cite{fang2018fair}.
Fig. \ref{f:intro} shows a typical application scenario of RBS system: smart home.
In Fig. \ref{f:intro}, the ceiling light is a RBS-equipped bulb with a RBS transmitter. Within the coverage of the RBS transmitter, all electronic devices embedded with the RBS receiver can be wirelessly charged at the same time.

Previous works \cite{zhang2018distributed, wang2019wireless} have already proposed methods to analyze the performance of RBS and answered the questions like ``how far RBS can reach" and ``how much power RBS can transfer". 
A multi-module distributed laser charging (DLC) system model is proposed in \cite{zhang2018distributed}  to illustrate the physical basis and mathematical formulas of RBS. On this basis,  the maximum power transmission efficiency is derived in a closed form \cite{zhang2018distributed}.
In \cite{wang2019wireless}, the consistent and stable operating conditions of the RBS system can determine the maximum power transmission distance and the power transmission efficiency within the working distance.
However, these analysis are applied after the resonant cavity is stabilized and have not explored the change of output power during the establishment process  of the resonant cavity.
The changing process of the resonant beam from non-steady state to steady state is still a mystery.
In other words, a complete and detailed transient analysis method is urgently needed for the RBS system.
Thus, we adopt equivalent circuit, a mature and effective method, to help analyze the transient characteristics of resonant cavity and simplify the calculation.
In fact, the equivalent circuit is another expression of the rate equation.
We verify the rationality of the equivalent circuit theoretically and complete the simulation experiment on it in Simulink.

\begin{figure}[t]
    \centering
    \includegraphics[width=0.85\linewidth]{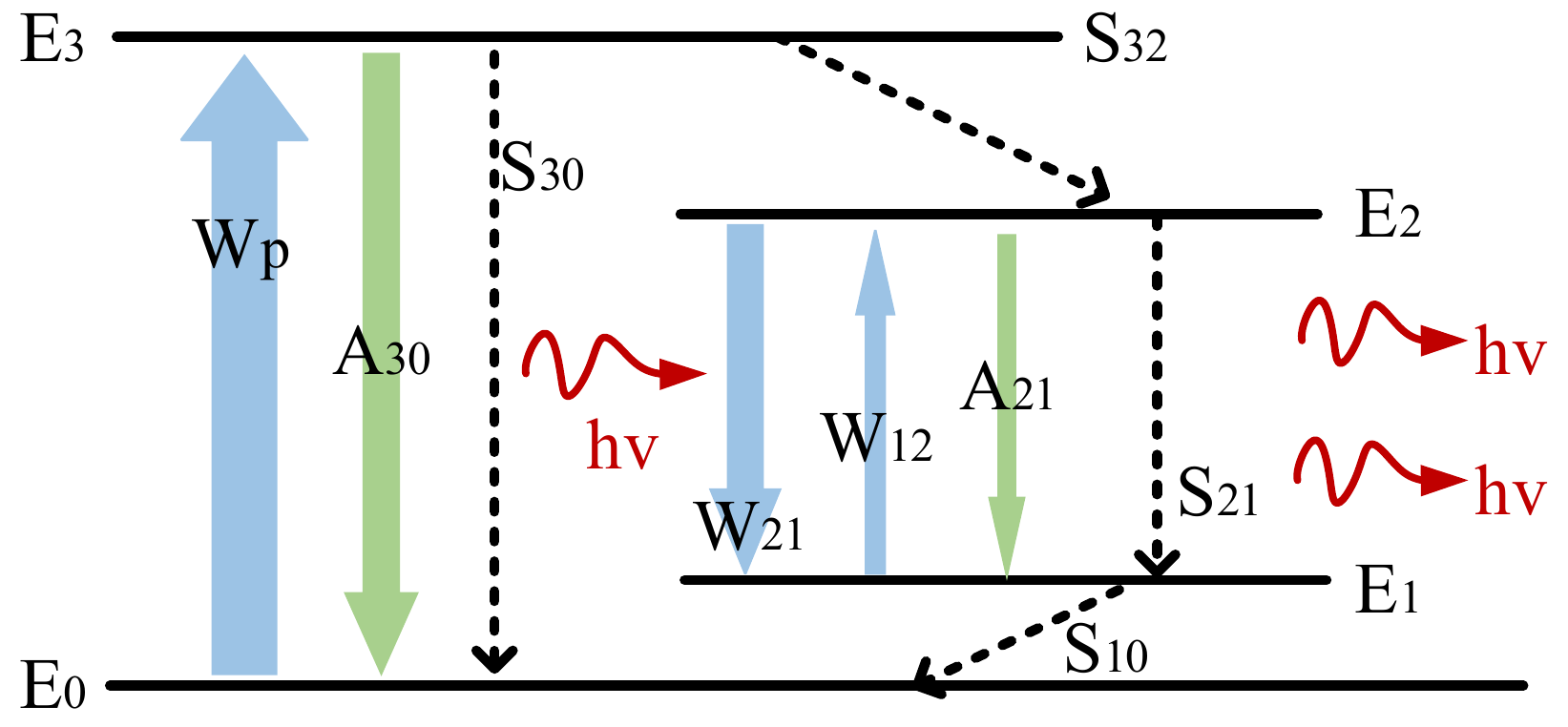}
    \caption{Particle transition and laser generation in a four-level system.}
    \label{f:fourlevel}
\end{figure}

In order to explore the transient change process of RBS establishment, we propose a resonant  beam  charging  and  communication system (RBCC) and an equivalent circuit analysis method.
The contributions of this paper are as follows.
\begin{enumerate}
\item We pay attention to the laser dynamic establishment process and its communication potential in RBS for the first time. We propose RBCC and a dynamic equivalent circuit analysis method, and verify its rationality in theory. The equivalent circuit explores the complete establishment process of the simultaneous wireless information and power transfer (SWIPT) system from scratch, and analyzes the change of output power during the establishment process.
\item The equivalent circuit has been well verified on the RBCC system, and the relaxation oscillation phenomenon of the RBCC system is analyzed for the first time. We further evaluate the charging and communication performance of the RBCC system through theoretical analysis and simulation testing.
\end{enumerate}

The remainder of this paper is as follows. 
Section \ref{sec:sys} outlines challenges facing SWIPT technology, and the basic models of a RBS system.
We introduce the mathematical principle and establishment process of the equivalent circuit in Section \ref{sec:design}.
Section \ref{sec:implet}  shows the simulation process of the equivalent circuit in simulink.
Performance of the proposed system is evaluated in Section \ref{sec:result} and conclusions are presented in Section \ref{sec:conclution}.

\section{PRELIMINARIES}\label{sec:sys}
In this section, we first briefly introduce SWIPT technology and its limitations. Then, we elaborate how RBS works. 

\subsection{SWIPT system}
Unlike traditional wireless communication that only transmits information, SWIPT can transmit energy to wireless devices while transmitting wireless signals \cite{perera2017simultaneous, 8759959}. 
The received energy is stored in the battery of the wireless device after a series of conversions \cite{boshkovska2015practical}.
The captured energy will be used for energy consumption of the normal information exchange circuit of the wireless device and the energy capture circuit. 
SWIPT technology is expected to achieve high-speed information exchange while efficiently feeding various terminal devices by extracting the energy in the received signal.
Thereby, SWIPT is able to mitigate the inconvenience caused by traditional wired or battery power supply. 
SWIPT reduces the size and cost of the terminal equipment, and greatly extends its standby time, which is particularly suitable for applications that require large-scale distribution of terminal nodes \cite{xu2018beam}.
SWIPT technology has broad application prospects in all aspects of life, such as smart home \cite{jang2018energy}, complex environments like oceans, forests, deserts \cite{zhang2019securing}, mines, and earthquake-stricken areas \cite{badirkhanli2020rescue}, and the field of biomedicine.

\begin{figure}[t]
    \centering
    \includegraphics[width=0.95\linewidth]{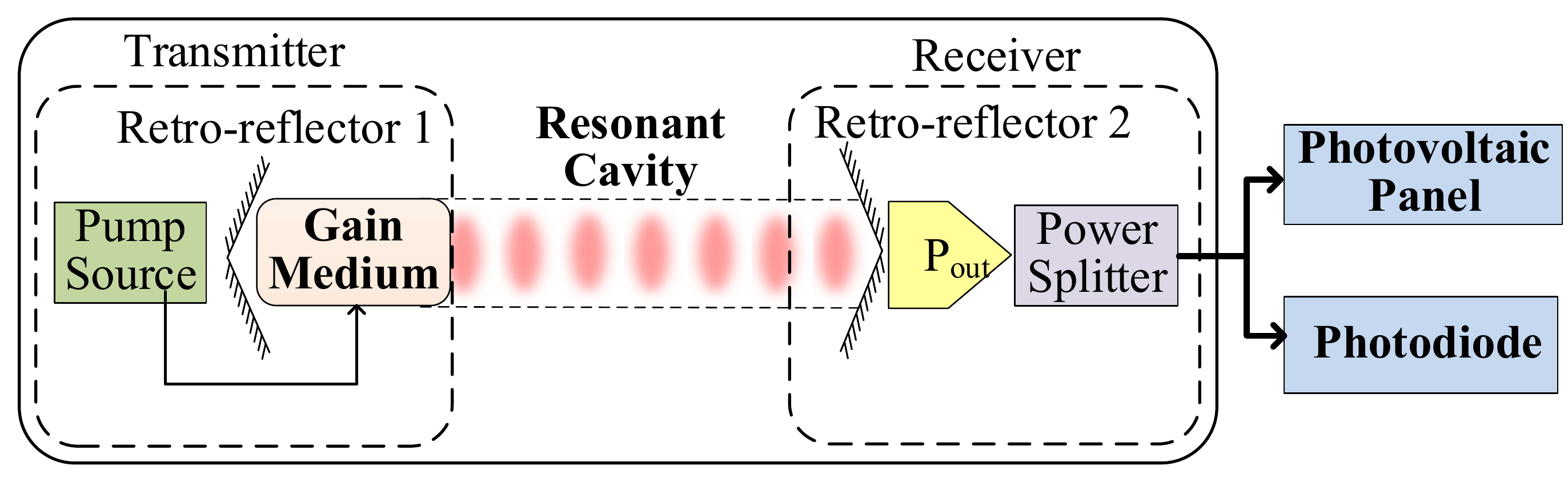}
    \caption{The overall structure of RBS.}
    \label{f:RBCStructure}
\end{figure}

However, SWIPT technology also faces many challenges.
Mobility will greatly affect the availability of channel state information.
Therefore, it is not easy to obtain accurate channel state information in the SWIPT system \cite{Huang2017simultaneous}.
Interference is one of the major challenges to mitigate in wireless communication \cite{hossain2019survey}.
In addition, security is also worthy of attention \cite{sun2019physical}.
As a typical SWIPT technology, RBCC has great potential in  overcoming the bottleneck of traditional SWIPT technology.

\begin{figure*}[htbp]
    \centering
    \includegraphics[width=0.9\linewidth]{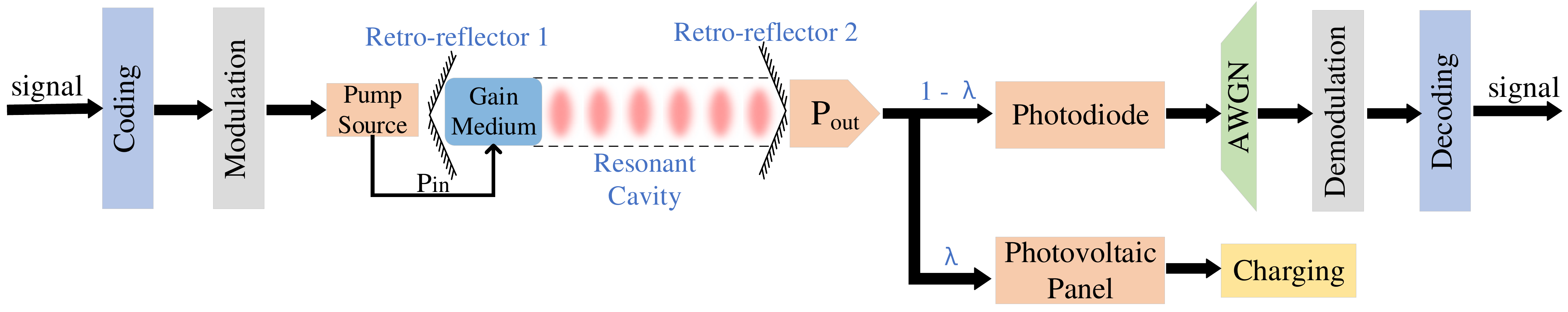}
    \caption{The communication and energy transmission model of the RBCC system.}
    \label{f:communication}
\end{figure*}

\subsection{RBS principle and structure}
RBS is essentially a wireless laser charging technology.
The core method of laser gain is stimulated radiation. 
When stimulated by the beam of light, 
the atoms in the crystal would transit from high energy levels to low ones, radiating light energy at the same frequency as the original light.

We choose Nd:YVO$_4$ as the material of gain medium crystal because  it is efficient and commonly used. 
Nd:YVO$_4$ exhibits pure four-level behavior due to its doped neodymium, as shown in Fig. \ref{f:fourlevel}. 
A four energy-level system achieves population inversion between two excited states $E_{\rm 2}$ and $E_{\rm 1}$.
Fig.~\ref{f:fourlevel} shows a four-level system where
$E_0$ is the ground state, $E_{\rm 1}$, $E_{\rm 2}$ and $E_{\rm 3}$ are excited states, and $E_{\rm 2}$ is the metastable state. 
$W_{\rm p}$ represents the rate of atom transitioning from the $E_{\rm 0}$ energy level to the $E_{\rm 3}$ energy level after receiving the excitation of the pump source. $A_{\rm 30}$ represents the rate that the atom will spontaneously radiate from the $E_{\rm 3}$ energy level to the $E_{\rm 0}$ energy level. $S_{\rm 30}$ represents the rate that the atom rapidly relaxes from the $E_{\rm 3}$ energy level to the $E_{\rm 0}$ energy level without radiation.
At the same time, the solid arrow represents the radiative transition, and the dashed arrow represents the non-radiative transition.
Therefore, the atom lifetime at the $E_{\rm 1}$ and $E_{\rm 3}$ energy levels is very short, while the atom lifetime at the $E_{\rm 2}$ energy level is long. Under the excitation of the pump source, the atoms on $E_0$ are excited to $E_{\rm 3}$, and then quickly transition to $E_{\rm 2}$. Because the atoms stay at the $E_{\rm 2}$ energy level for a long time, there are many atoms at the $E_{\rm 2}$ energy level. The atoms that transition from the metastable state $E_{\rm 2}$ to $E_{\rm 1}$ will quickly return to $E_0$, so the number of atoms on the $E_{\rm 1}$ energy level is small. In this way, a population inversion can be formed between the energy levels of $E_{\rm 1}$ and $E_{\rm 2}$.
Since the lower energy level $E_{\rm 1}$ of the four-level system to achieve population inversion is an excited state rather than a ground state, under normal circumstances, the number of atoms on it is already very small. Therefore, as long as there is a slight accumulation of atoms in the metastable state $E_{\rm 2}$, the population inversion can be easily achieved, which is a prerequisite for the lasing process.

From a macro perspective,
the whole RBS is divided into two separate parts, which are the transmitter and the receiver, as shown in Fig.~\ref{f:RBCStructure}. 
In the RBS transmitter, there are a retro-reflector 1 ($\rm RR1$) with 100\% reflectivity and a gain medium, which is used to amplify photons. A retro-reflector 2 ($\rm RR2$) with 80\% reflectivity and a power splitter compose the RBC receiver. 
In the whole process, pump source is activated by current, and provides all energy the RBC system needs to start up and keep stable operation. 
The optical power converted from electrical power by the pump source oscillates back and forth between $\rm RR1$ and $\rm RR2$ after amplified by the gain medium in transmitter. 
Besides, there are several losses in the RBS consuming energy, such as air loss, mirror transmission in $\rm RR2$, scattering and so on. A steady resonant beam forms only after gain and loss are in balance.
Then the receiver of RBS could output stable power $P_{\rm out}$, which is divided into two parts by power splitter and used to charge the photovoltaic panel as well as communicating with photodiode.

RBS is a novel wireless energy transmission technology with high power and long distance. 
Based on its unique architecture, RBS achieves mobile charging with automatic alignment and intrinsic safety. 
Retro-reflectors like $\rm RR1$ and $\rm RR2$ are able to make the incident light reflect back along the original path. In this condition, photons generated by pump source  can oscillate back and forth between the transmitter and the receiver thereby form a stable resonant beam, regardless of the position of the receiver.
As for intrinsic safety, resonance will stop immediately as long as one obstacle blocks resonant beam path, so that the obstacle like person's hand or other fragile things will not be hurt. 
When an obstacle enters the transmission path, losses in the cavity increase rapidly and destroy the balance between gain and losses. Resonant beam fades away immediately so that safety can be guaranteed. On the contrary, when the obstacle leaves the transmission path, the cavity loss becomes smaller, so the resonance will automatically re-establish.

\section{Transient Analysis}\label{sec:design}
In order to make a complete and unified analysis of the RBCC performance, we add the communication part to the previous energy transmission system diagram, as shown in Fig.~\ref{f:communication}. The electrical signal, which contains the information to be transmitted, is coded and modulated before being used to excite the pump. 
The electrical signal is pumped into an optical signal and then exits $\rm RR2$ through the resonant cavity. 
The output optical signal is divided into two parts, and converted into electrical signals through the photodiode (PD) and electrical power through photovoltaic (PV) panel, respectively.
One part is used to charge the device. In order to simulate the real communication channel, the other part is added with noise, and demodulated and decoded to obtain the original signal.

\begin{figure}[t]
    \centering
        \includegraphics[width=3.2in]{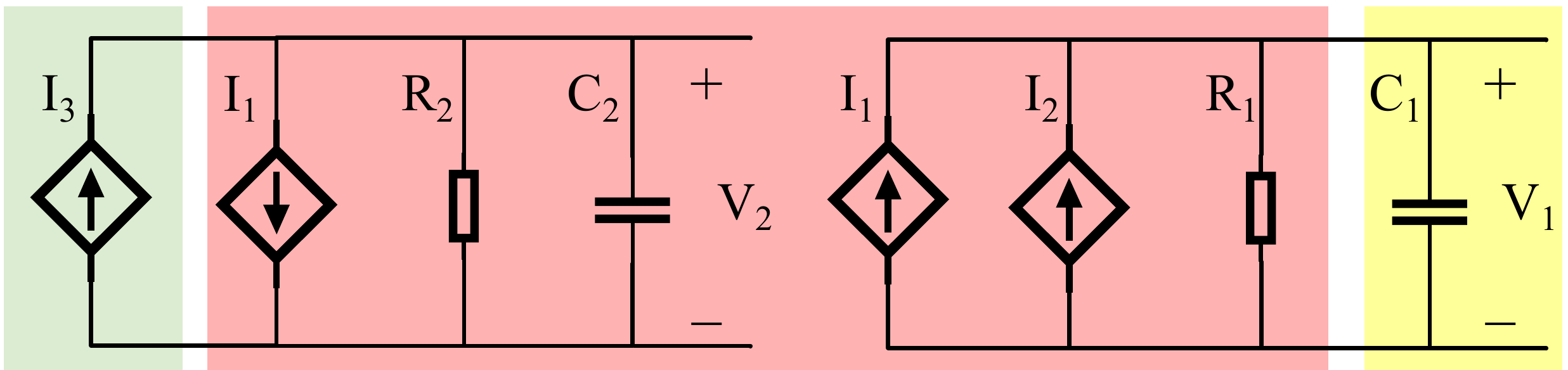}\\
    \caption{The schematic diagram of the four-level equivalent circuit.}
    \label{f:circuit}
\end{figure}

The method of constructing its circuit model with the rate equation of optoelectronic devices is often used to simulate and analyze the DC, AC and transient response characteristics of photoelectric devices such as lasers and detectors \cite{yokoyama1989rate}. 
By analyzing the optical characteristics and rate equations of the gain medium, we construct an equivalent circuit to simulate the RBCC system and dynamically analyze its establishment process. In addition, we have also established a complete communication system to achieve stable and reliable communication.

\subsection{Equivalent circuit}
The transient characteristics of semiconductor lasers are determined by two rate equations, namely the photon rate equation and the carrier rate equation.
The rate equation is another expression of the laser gain process.
From \cite{yong2013equivalent}, we have
\begin{equation}
  \frac{\partial \Phi}{\partial t} = c \Phi \sigma n - \frac{\Phi}{\tau_c} + S,
  \label{e:rateEquation1}
\end{equation} 
\begin{equation}
  \frac{\partial n_2}{\partial t} = -c \Phi \sigma n_2 - \frac{n_2}{\tau_f} + W_p n_0,
  \label{e:rateEquation2}
\end{equation} 
where $\Phi$ is photon density in resonant cavity;
$c$ is speed of light in the medium; $\sigma$ is stimulated emission cross section; $S$ can be expressed as the rate at which spontaneous emission contributes to stimulated emission; 
$n$ is the inversion population density, i.e., the density difference between high-energy and low-energy particles;
$\tau_c$ and $\tau_f$ are the decay time for photons in the optical resonator and fluorescence decay time, respectively. 
In equation~(\ref{e:rateEquation2}), $n_2$ and $n_0$ are the particle density above the corresponding energy level;
$W_p$ is pumping rate. In reality, $n_0 \approx n_2 $.

The equivalent circuit equation obtained by normalizing and rearranging the rate equation is as follows:
\begin{equation}
  C_1\frac{\partial V_1}{\partial t} +  \frac{V_1}{R_1}= I_1  + I_2,
  \label{e:rate1}
\end{equation} 
\begin{equation}
  C_2\frac{\partial V_2}{\partial t} + \frac{V_2}{R_2}= I_3 - I_1, 
  \label{e:rate2}
\end{equation} 
where $C_1 = C_2 = 1$; $R_1 = \tau_c$; $R_2 = \tau_f$; $I_1 = c \Phi \sigma n_2$; $I_2 = S$; $I_3 = W_p n_0 $; $V_1$ is the photon density in the resonant cavity and $V_2$ represents particle density of energy level $E_2$, which is a metastable state. 
Here we also give the calculation formula of $I_3$, as shown below:
\begin{equation}
  I_3 = W_p n_0 \\  =  \frac{P_{\rm in}}{v_L V},
  \label{e:G3}
\end{equation} 
where $P_{\rm in}$ is input power, $v_L$ is incident laser frequency and $V$ is the volume of gain medium.

If $I_1$, $I_2$, and $I_3$ are regarded as controlled current sources, $V_1$ and $V_1$ are regarded as voltages, $R_2$ and $R_3$ are regarded as resistors, and $C_2$ and $C_3$ are regarded as capacitors, then equations (\ref{e:rate1}) and (\ref{e:rate2}) can be interpreted as Kirchoff’s current law equations. The equivalent circuit representation of the rate equation is shown in Fig. \ref{f:circuit}.
$V_2$ is the voltage across capacitor $C_2$ in circuit on the left.
As a controlled current source, $I_1$ is controlled by $V_2$. So it takes circuit on the left some time to establish stability instead of becoming stable immediately. This phenomenon will lead to relaxation oscillations.
At the same time, $I_1$ in circuit (b) is also controlled by $V_2$, thus influences the establishment process of circuit steady state. 
For output light power $P_l$, we have
\begin{equation}
  P_l = V_1 h v_L,
  \label{e:outputPower}
\end{equation} 
where $h$ is the Planck constant and $v_L$ is frequency of light.

On this basis, we introduce equivalent circuit of the four-level energy system to simulate the build process of intracavity resonant beam so that we are able to analyze the dynamic  RBCC system.
We illustrate the four-level equivalent circuit and its correspondence with RBCC system in Fig.~\ref{f:circuit}. 
Fig.~\ref{f:circuit}  constitutes a complete equivalent circuit and there are energy interaction between them. 
Equivalent circuit has a clear and understandable structure and consists of controlled current sources, resistors and capacitors.
The green shadow part in Fig.~\ref{f:circuit} is the input power of equivalent circuit and it corresponds to pump source of RBCC system in Fig. \ref{f:RBCStructure}. 
The yellow shadow part in Fig. \ref{f:circuit} represents output power. 
As for the remaining red part, it corresponds to RBC spatially separated cavity and could be used as a power transmission and communication channel.

We use equivalent circuit to dissect build process of resonant beam and fill the gap between static analysis and dynamic analysis. Equivalent circuit simulates the process of how photons become a stable resonant beam and intuitively reflects how the change of pump source and losses in the channel alter output power and influence the performance of communication and energy transmission.

\subsection{Energy conversion on the pump source}
In the whole system, the optical signal and the electrical signal undergo two mutual conversions, respectively, on the pump source and photoelectric conversion devices, including a PV panel and a PD.
After coding and modulation, input electrical signal is converted to optical signal through pump source. 
Then, the output optical power $P_{\rm pump}$ of the pump can be described as \cite{liu2005semiconductor}
\begin{equation}
  P_{\mathrm{pump}}\left(I_{\mathrm{in}}\right)=\frac{h c}{q \lambda} \eta_{\mathrm{e}}\left[I_{\mathrm{in}}-I_{\mathrm{th}}\right],
  \label{e:P_out}
\end{equation}
and
\begin{equation}
  \eta_{\mathrm{e}}=\eta_{\mathrm{inj}} \frac{\gamma_{\mathrm{out}}}{\gamma_{\mathrm{c}}},
  \label{e:P_rta}
\end{equation}
where $h$ is Planck’s constant; $c$ is the speed of light in
vacuum; $q$ is the electron charge; $\lambda$ is the emission wavelength;
$\eta_{\rm e}$ is the external quantum efficiency;  $I_{\rm th}$ is the constant
threshold current and $I_{\rm in}$ is the input current. The carrier injection efficiency $\eta_{\rm inj}$ and $I_{\rm th}$ are temperature-dependent parameters. $\gamma_{\rm out} / \gamma_{\rm c}$ is the photon extraction efficiency.

\begin{figure}[t]
    \centering
    \includegraphics[width=3in]{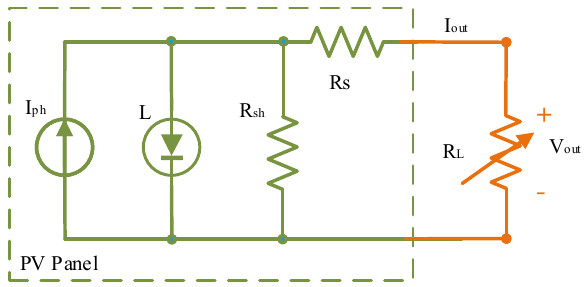}
    \caption{The circuit principle of photovoltaic panels.}
    \label{f:PV}
\end{figure}

\begin{table}[t]
  \centering
  \caption{Parameters in the simulation setting \cite{yong2013equivalent}.}
    \begin{tabular}{cc}
      \toprule  
      PARAMETER & VALUE \\  \hline  
      $\tau_c$, all the losses in an optical \\ resonator  of a laser oscillator & $4.4 \times 10^{-4}$s \\  \hline  
      $\tau_f$, fluorescence lifetime of energy \\ level  on gain medium & $2.3 \times 10^{-4}$s \\  \hline 
      $c$, speed of light in the medium  & $1.67 \times 10^8$ m/s \\  \hline  
      $\delta$, stimulated emission cross section & $2.8 \times 10^{-19}$ cm$^2$ \\  \hline  
      $S$, spontaneous emission of photons \\ coupled into the cavity & / \\  \hline  
      $v_L$, laser frequency & $2.82 \times 10^{14}$ Hz \\  \hline 
      $h$, planck constant & $6.63 \times 10^{-34}$ m$^2$kg/s\\ 
      \bottomrule
    \end{tabular}
  \label{t:parameter}
\end{table}

\subsection{Photoelectric conversion for charging}

An ideal PV panel can be modeled as a photocurrent source $I_{\rm ph}$ connected in parallel with a diode with a forward current of $I_{\rm L}$, which is the current flowing through the diode. 
Fig. \ref{f:PV} shows the equivalent circuit of PV panels, which also includes a parallel resistor $R_{\rm sh}$ and a series resistor $R_{\rm s}$ as used in~\cite{chaibi2018new, lineykin2014improved}.
Therefore, according to Kirchhoff's law, the current-voltage (I-V) characteristics of PV panels can be described as \cite{rahman2014generalised,rustemli2011modeling}:
\begin{equation}
I_{\mathrm{out}} =I_{\mathrm{ph}}-I_{\mathrm{L}}-\frac{V_{\mathrm{L}}}{R_{\mathrm{sh}}},
\end{equation}
and 
\begin{equation}
I_{\mathrm{L}} =I_{0}\left(\mathrm{e}^{\frac{V_{\mathrm{d}}}{n_{\mathrm{s}} n V_{T}}}-1\right),
\end{equation}
\begin{equation}
V_{\mathrm{L}} =V_{\mathrm{out}}+I_{\mathrm{out}} R_{\mathrm{s}},
\end{equation}
where $I_0$ is the reverse saturation current and $n_s$ is the number of batteries connected in series in the PV panel. Therefore, for single-cell PV panels, $n_s = 1$. In addition, $V_{\rm T}$ is the temperature-dependent thermal voltage of the diode junction.

The photocurrent $I_{\rm ph}$ is related to the light power on the
PV panel, and 
can be estimated by \cite{hasan2016overview}:
\begin{equation}
    I_{\mathrm{ph}} =\rho_1 P_{\mathrm{E}} \\ =\rho\left\{\eta_{\mathrm{s}} f(d)\left\{\frac{h c}{q \lambda} \eta_{\mathrm{e}}\left[I_{\mathrm{in}}-I_{\mathrm{th}}\right]\right\}+C\right\}, 
    \label{e:pv}
\end{equation}
where $\rho_1$ is the optical-to-electrical conversion responsivity in A/W and can be measured under the standard test condition (STC, 25 $^{\circ}$C temperature and 1000 W/m$^2$ irradiance).

\subsection{Communication Channel}
To present the signal gain in the transmission channel of the RBCC, we have signal power output from PD as~\cite{smestad2004conversion}
\begin{equation}
    I_{\mathrm{pd}} =\rho_2 P_{\mathrm{C}}, 
    \label{e:pd}
\end{equation}
where $\rho_2$ represents the optical-to-electrical conversion
responsivity of PD, and $P_{\mathrm{C}}$ is the power used for communication, which is one part of the  optical power output by the RBCC.
To evaluate the performance of the communication in the RBCC system,
we analyze the signal-to-noise ratio (SNR) of the channel. From~\cite{demir2017handover}, we have 
\begin{equation}
    {\rm SNR} = \frac{I_{\mathrm{pd}}^2}{\sigma_{\rm noise}},
    \label{e:SNR}
\end{equation}
where $\sigma_{\rm noise}$ is the variance of the noise in the channel.

In the RBCC system, 
the signals in $P_{\rm C}$ will be demodulated and decoded with noises and output electrical signals $S_{\rm o}$. 
Actually, every section in communication will introduce different noises. We eventuate these noises into one section and use additive white Gaussian noise (AWGN) to replace them, for the purpose of easy understanding and calculation. 
Eventually, we get $S_{\rm o}$ and information is passed.

\subsection{Trade-off between energy charging and communication}
As a SWIPT system, RBCC performs wireless charging and wireless communication at the same time. So the optical power output by the RBCC  is divided into two parts at the receiver. One part is used for communication, and the other part is used for energy charging, named $P_{\rm C}$ and $P_{\rm E}$, respectively.
Then, $P_{\rm C}$ is converted by PD; and $P_{\rm E}$ is converted into electrical power through PV panel.
The charging power $P_{\rm E}$ will be stored in the battery to maintain the normal operation of the device and the subsequent processing of communication signals~\cite{8792959}.

The communication performance of RBCC is directly correlated with $P_{\rm C}$. By controlling the energy allocation between charging and communication, it is possible to provide high charging efficiency while ensuring communication quality.

\begin{figure*}[t]
    \centering
    \includegraphics[width=6.4in]{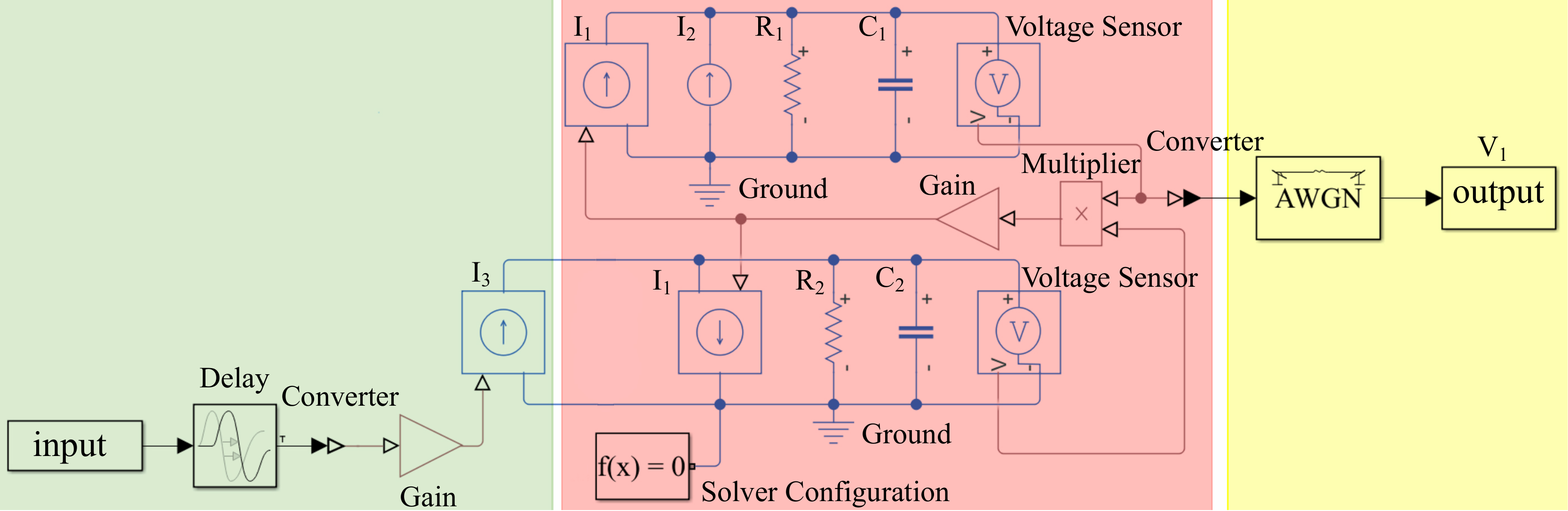}
    \caption{Equivalent circuit in simulink.}
    \label{f:circuit_system}
\end{figure*}

\section{Implementation}\label{sec:implet}
Table \ref{t:parameter} provides all the parameters of the rate equation.
All simulations are carried out with Simulink~\cite{salmi2012matlab}. Current source is used to drive the laser in order to compare the simulated results with analytical and numerical results.

\subsection{Simulation model}
We establish a mapping relationship between the output voltage of circuit and the optical power so that we can reveal the whole establishment process of resonance light from simulated results.
We complete circuit model in Simulink, as shown in Fig. \ref{f:circuit_system}, and apply a certain bias current to the gain medium to make it work in linear mode, which is included in $I_{\rm 3}$.
In order to reduce the influence of relaxation oscillation on the communication performance of the entire circuit, we have added a delay module to the signal input. 
The output of the whole system is $V_{\rm 1}$, which first appeared in Fig.$\ref{f:circuit}$, and will be converted to light power in subsequent processing.
Limited by rules of Simulink, we add some computing devices in the circuit, such as gains, multiplier, converters, voltage sensors, solver configuration and so on. 
Converters and voltage sensors are used for conversion between different types of signals.
Solver configuration is used to set the simulation accuracy and other parameters.
These devices just help achieving function of circuit and will not make any impact to final results.

To test and verify the communication performance of RBCC, we imitate Fig. \ref{f:communication} and add some modules to the original basic model, as shown in the shaded part in Fig. \ref{f:circuit_system}.
We load the coded and modulated signal and input it into the circuit after a certain delay. The delay here is set for the purpose of avoiding relaxation oscillations, which is observed from the output of Fig. \ref{f:circuit_system}.

\subsection{Energy distribution}
As a SWIPT system, part of the power output by the circuit is used for charging and the other part is used for communication. In the experiment, we set a power splitting ratio $\lambda \in [0,1]$, where $\lambda = P_{\rm charge} / P_{\rm total}$, as shown in the Fig. \ref{f:communication}. 
$\lambda = 1$ means that the output power is all used for charging, while $\lambda = 0$ means that all output power is used for communication. By adjusting the value of $\lambda$, we can get any value of charging or communication power.
As in~\cite{9435099}, we use a beam splitter to realize the power split of the beam.

\section{Result}\label{sec:result}
In this section, we first introduce relaxation oscillation which is observed by the equivalent circuit. 
Then, we analyze the influence of pump power on SWIPT in detail, and give the corresponding resonant cavity design scheme.
Finally, we introduce the communication performance of the RBCC system in detail.

\subsection{Relaxation oscillation}
\begin{figure}[t]
    \centering
    \includegraphics[width=3.5in]{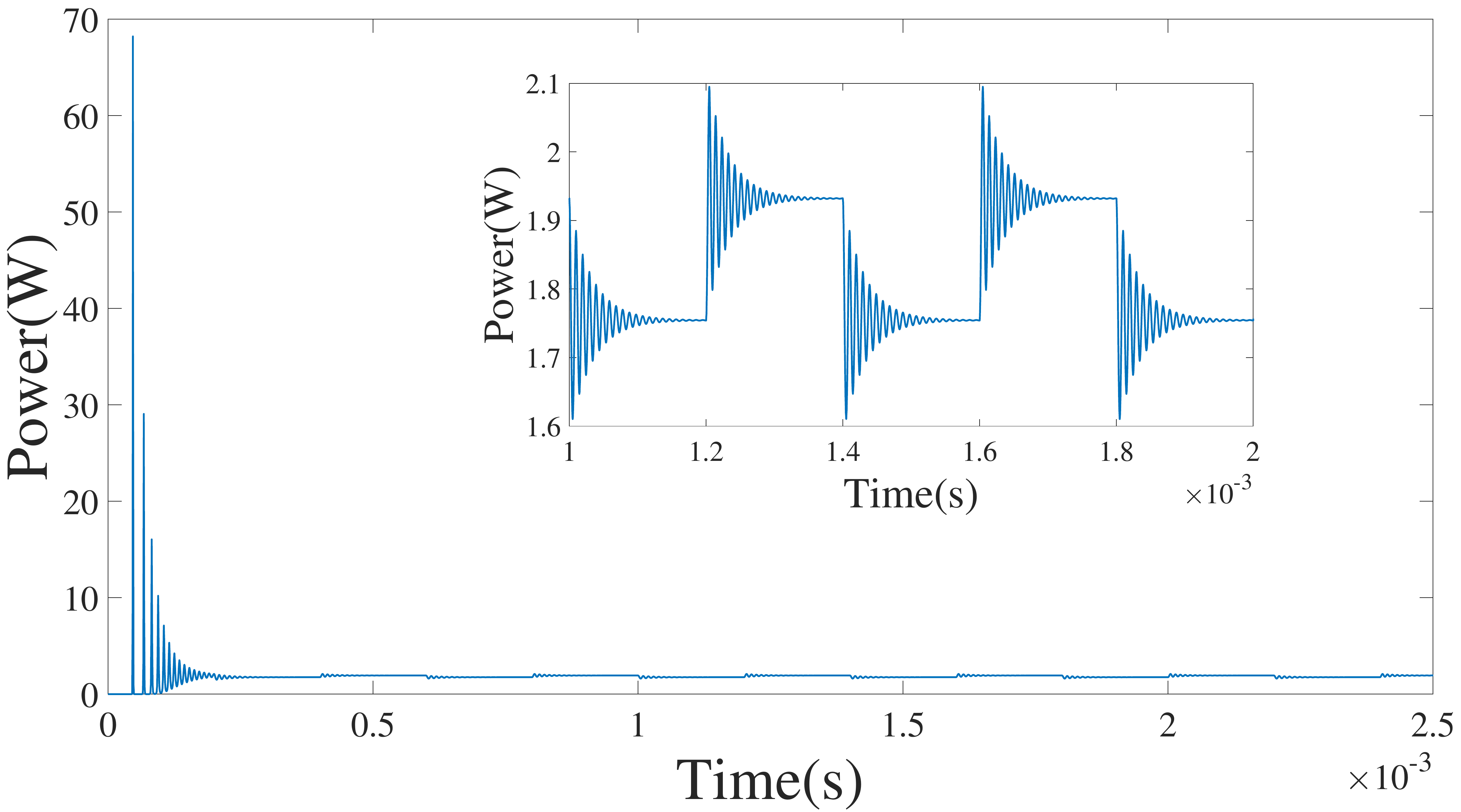}
    \caption{Relaxation oscillation phenomenon in output power.}
    \label{f:Relaxation}
\end{figure}

The equivalent circuit provides an excellent theoretical support and practical guidance for exploring the dynamic characteristics of information and energy simultaneous transmission systems, such as RBCC systems. We establish the simulation shown in Fig. \ref{f:circuit_system} in simulink, and connect the simulation with the actual model. When the pump source can drive the entire system, the RBCC system will have relaxation oscillations before establishing a stable charging beam, as shown in Fig. \ref{f:Relaxation}. The sub-figure in Fig. \ref{f:Relaxation} illustrates that each time the balance is broken and the system re-establishes a stable beam, there will be relaxation oscillations in the output power.

We observe that there is no output power at the beginning. After the populations of $E_1$ and $E_2$ in the gain medium are reversed, the resonant cavity starts to generate laser light, and thus has optical power output. However, the optical power at this time is not stable, and a relaxation oscillation phenomenon occurs. 
In Fig. \ref{f:Relaxation}, we applied 30 W of bias and 0.3 W of signal power to the pump source. Here we use an on-off keying (OOK) modulation scheme.
In this case, the relaxation oscillation can be as high as 68 W. 
This is caused by changes in the relative relationship of the pump and inversion populations. The phenomenon of relaxation oscillation also reminds us to pay attention to the highest withstand voltage value of components in related circuits, so as to avoid breakdown due to excessive instantaneous power.

Since the input pump power is a periodically changing square wave signal, the corresponding waveform changes can also be observed in the output power. This shows that the RBCC system has great communication potential and can transmit information while transmitting energy wirelessly. The sub-figure of Fig. \ref{f:Relaxation} shows the details of the corresponding change in the output power at the moment of the square wave signal change. It can be seen that at every moment when the input waveform changes, the output waveform needs to re-establish a stable state. The percentage of the settling time in the duration of a single waveform is determined by the frequency and power of the input waveform. So when we use RBCC for communication, this is a very interesting feature and 
a noteworthy aspect.

\subsection{Relaxation oscillation for SWIPT system}
\begin{figure}[t]
    \centering
    \includegraphics[width=3.5in]{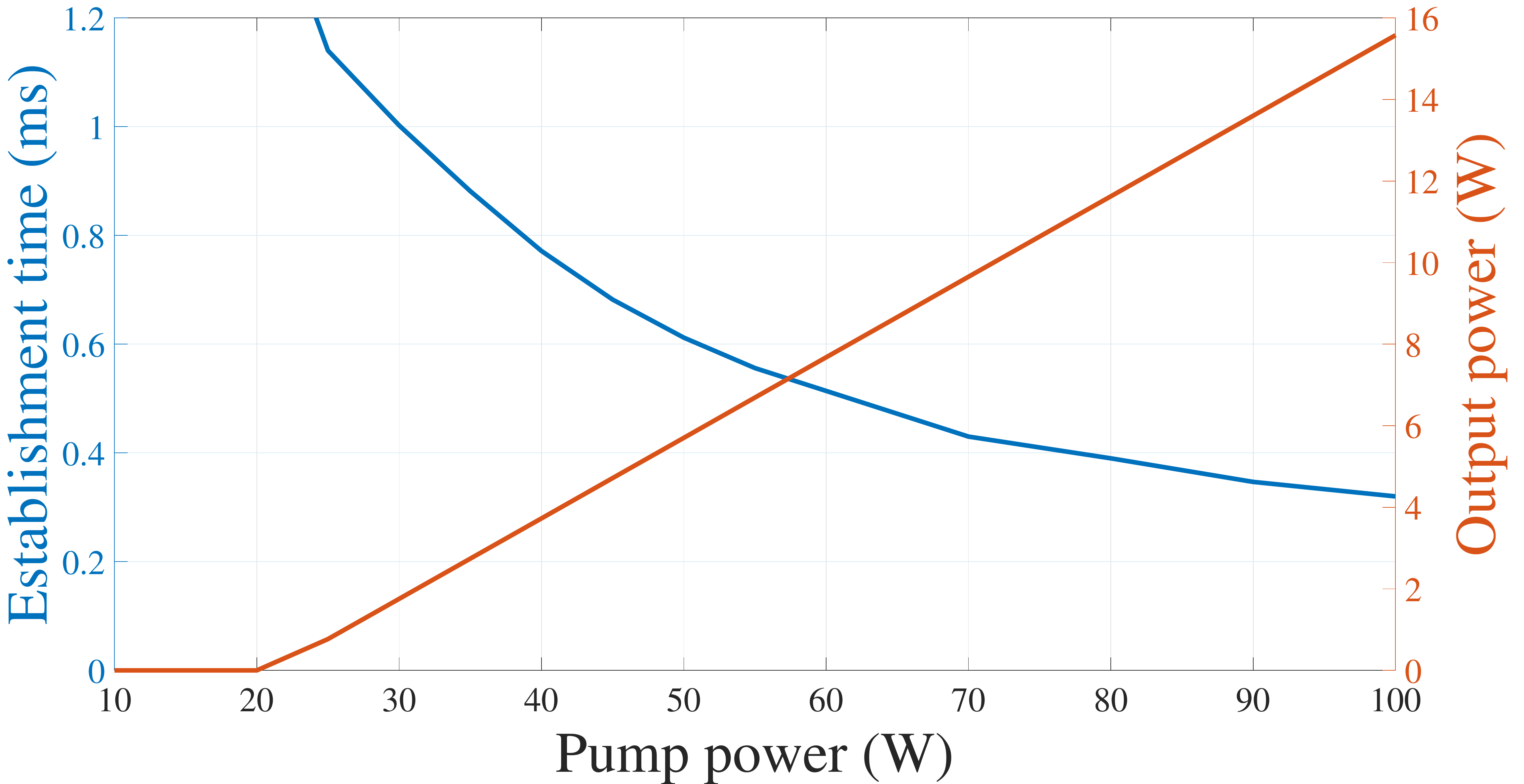}
    \caption{The relaxation oscillation and output power vary with the pump power.}
    \label{f:changePowert}
\end{figure}

Fig. \ref{f:changePowert} shows how the establishment time of relaxation oscillation and total light power output outside the cavity vary with the pump power increases. 
As the pump power increases, the output power increases, but the settling time of the relaxation oscillation gradually decreases. That is to say, the greater the pump power, the faster the entire system will enter a stable state.
When the pump power is less than 20 W, the output power is basically zero. At this time, the pump power is too small so that a stable resonant beam cannot be formed in the cavity, and the establishment time of relaxation oscillation is infinite.
Experiments show that the 25 W pump power is the minimum pump power that can make the system output a stable optical power. 
From Fig. \ref{f:changePowert}, we can see that as the pump power continues to increase, the output power continues to increase, and the power conversion rate is also increasing, from 5.85\% to 15.58\%.
This is an important discovery for energy transmission. As a wireless charging system for RBCC, energy utilization is a performance indicator that needs to be considered first. With the relationship between pump power and output power, we can select the appropriate pump power to drive the entire system.
At the same time, the duration of the relaxation oscillation is continuously decreasing, from 1.14 milliseconds to 0.32 milliseconds. 
This phenomenon is also in line with the previous theoretical argument. 
From this we can know that when using RBCC for communication, the influence of the duration of the relaxation oscillation should be considered.

\subsection{Communication performance of RBCC}
\begin{figure}[t]
    \centering
    \includegraphics[width=3.5in]{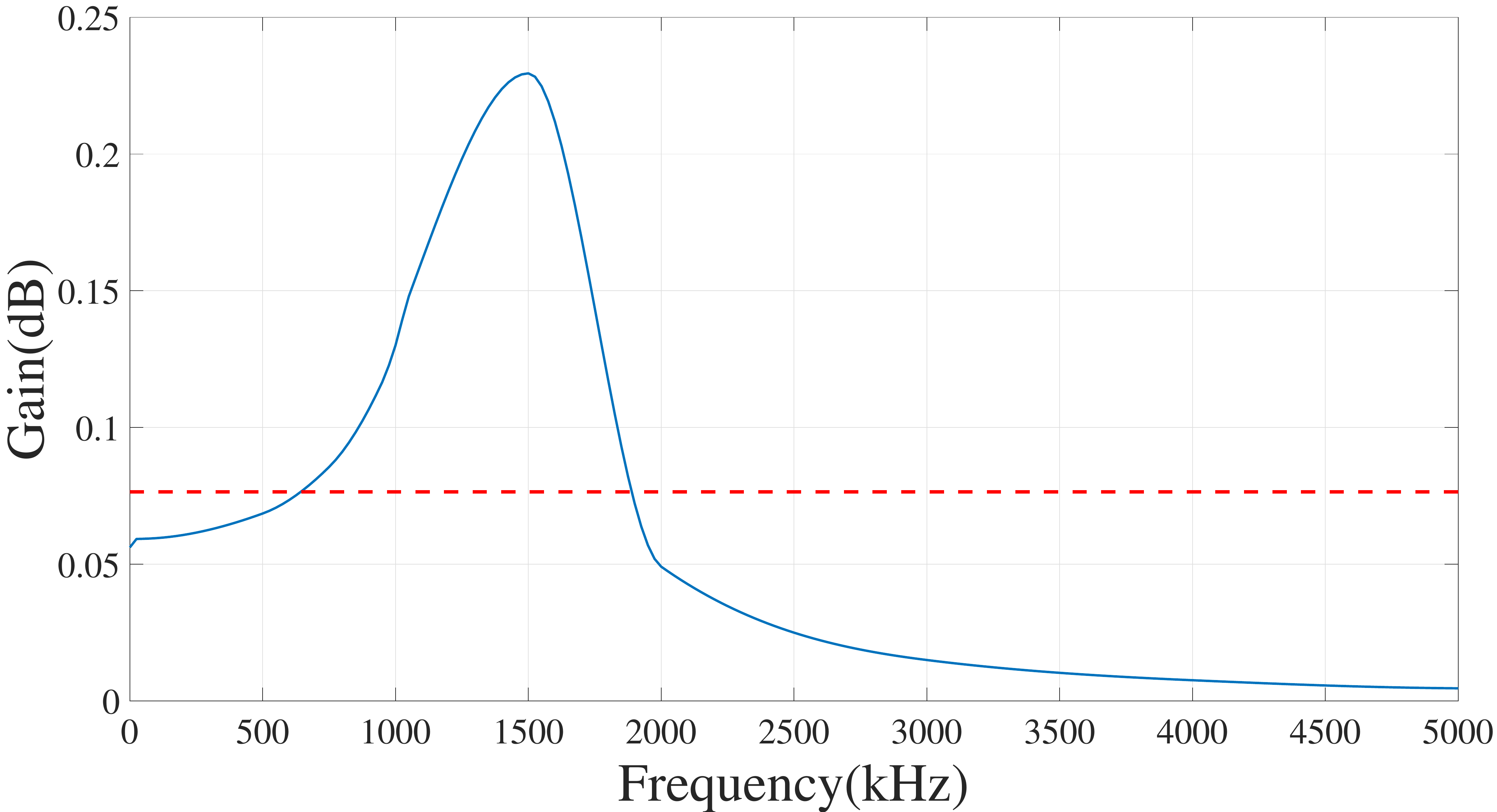}
    \caption{The spectrum response of the RBCC channel. }
    \label{f:bandwidth}
\end{figure}

We study the communication performance of the RBCC system in depth and show the channel frequency response in Fig. \ref{f:bandwidth}, 
and the input signal is a sinusoid.
While keeping the input power constant, we test the response of the system to inputs of different frequencies.
The red dotted line indicates that gain becomes 0.5 times of the maximum value.
The bandwidth of the RBCC channel is about 1 MHz.

\begin{figure}[t]
    \centering
    \includegraphics[width=3.5in]{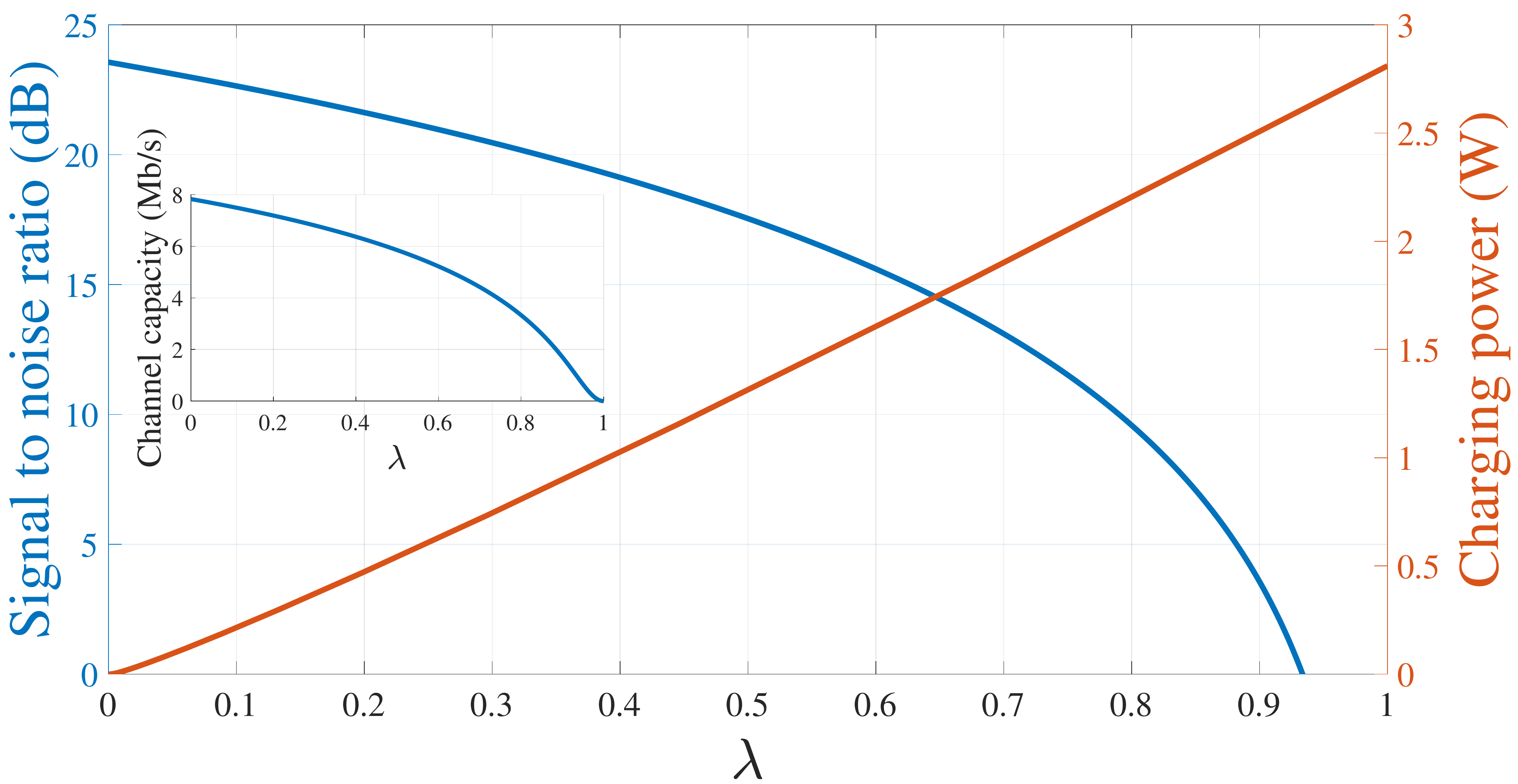}
    \caption{The effect of $\lambda$ on the signal-to-noise ratio, channel capacity, and output power with Gaussian white noise.}
    \label{f:lambda}
\end{figure}

As a SWIPT system, RBCC can allocate energy for power transmission and communication by adjusting the power splitting ratio $\lambda$. This makes the communication performance of the RBCC system closely related to $\lambda$. 
Fig. {\ref{f:lambda}} shows the change trend of SNR, channel capacity, and charging power with different $\lambda$. 
The change trend of channel capacity is basically proportional to SNR and inversely proportional to charging power.
When $\lambda$ is 0, all output power is used for communication, at this time the SNR reaches a maximum of 23.54 dB and channel capacity is about 8 Mb/s. As $\lambda$ gradually increases, the SNR decreases gradually and varies sharply when $\lambda$ is around 0.7. The change trend of the SNR also confirms to the mathematical definition of the SNR. When $\lambda$ is near 0.93, the SNR is 0 dB. In this process, the output power increases linearly as a function of  $\lambda$.

\begin{figure}[t]
    \centering
    \includegraphics[width=3in]{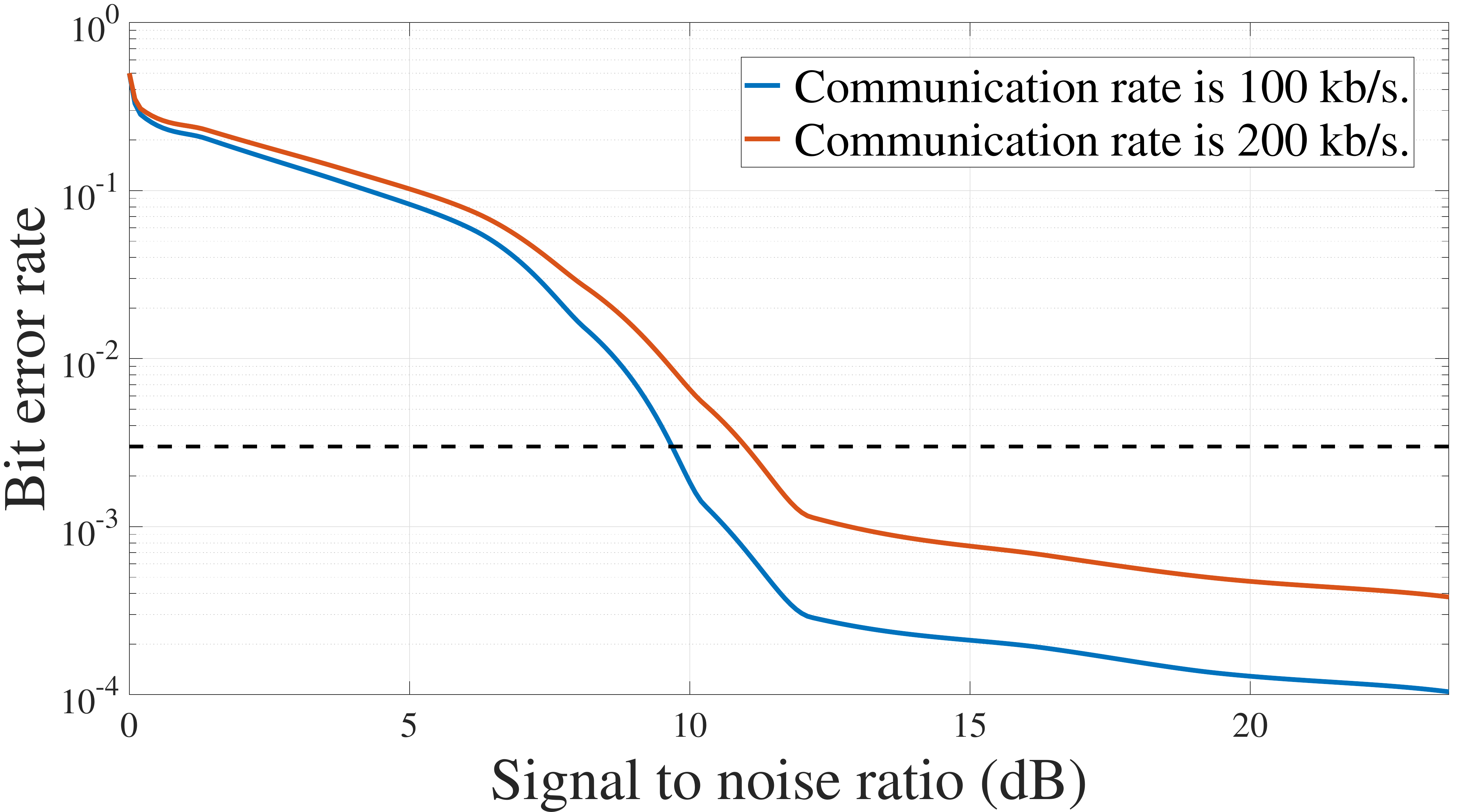}
    \caption{Under different communication rates, the bit error rate changes with the signal-to-noise ratio.}
    \label{f:SER}
\end{figure}

Fig. {\ref{f:SER}} shows the variation of the system error rate with the SNR at different communication rates.
For Fig. {\ref{f:SER}}, the signal is OOK modulated. The blue solid line in the figure corresponds to the case when 
communication rate is 200 kb/s, and the orange solid line indicates that the communication rate is 100 kb/s. The black dashed line represents a reference line parallel to the x-axis with a value of 3 $\times 10^{-3}$. In other words, only if the bit error rate is less than 3 $\times 10^{-3}$, the communication can proceed normally. 
When other conditions are the same, the low communication rate obviously has a lower bit error rate than the high communication rate. In other words, there is a tradeoff between the communication rate and the bit error rate. When the SNR approaches 0 dB, the signal-to-noise ratio approaches 50\%. When the SNR reaches the maximum, the system has the lowest bit error rate of  $3.83 \times 10^{-4}$ with 200 kb/s communication rate, and the lowest bit error rate of  $ 1.04 \times 10^{-4}$ with 100 kb/s communication rate.

\section{Conclusion}\label{sec:conclution}
This paper proposed RBCC, a new SWIPT system, and used equivalent circuit analysis method to conduct transient analysis on it.
Starting from the rate equation, we deduced the mathematical expression of the equivalent circuit.
Then, we realized the simulation of equivalent circuit in simulink and simulated the entire process of the communication channel. We found the relaxation oscillation phenomenon of the RBCC system before stabilization through the change of the circuit output power, and made a complete analysis of the relaxation oscillation. Finally, we analyzed the energy transmission performance and communication performance of RBCC.


%





\ifCLASSOPTIONcaptionsoff
  \newpage
\fi



%

\bibliographystyle{IEEEtran}
\bibliography{Reference}

\end{document}